\documentclass{aa}  
\usepackage{graphicx}
\usepackage{txfonts}
\usepackage{natbib}
\bibpunct{(}{)}{;}{a}{}{,}
\begin{document}
   \title{A spectroscopic study of the Open Cluster NGC~6475 (M~7)}

   \subtitle{Chemical Abundances from stars in the range T$_{\rm eff}$= 4500-10000 K}

   \author{Sandro Villanova
          \inst{1}
          \and
          Giovanni Carraro
          \inst{2}
          \and
          Ivo Saviane
          \inst{2}
          }

   \offprints{}

   \institute{Grupo de Astronomía, Departamento de Física, Casilla 160, 
              Universidad de Concepción, Chile\\
              \email{svillanova@astro-udec.cl}
         \and
              ESO, Alonso de Cordova, 3107 Vitacura, Santiago de Chile, Chile\\
              \email{gcarraro@eso.org,isaviane@eso.org}
              }

   \date{Received .....; accepted .....}

% \abstract{}{}{}{}{} 
% 5 {} token are mandatoryms_Dec20.tex
 
  \abstract
   {}
   {Clusters of stars are key objects for the study of the dynamical
    and chemical evolution of the Galaxy and its neighbors. In particular chemical 
    composition is obtained from different kinds of stars (hot main-sequence
    stars, cool main-sequence stars, horizontal-branch stars, RGB stars) using different
    methodologies. Our aim is to apply these methodologies to the stars
    of the Open Cluster NGC~6475. Obtaining a census of the most 
    important elements we will be able to test their consistence.
    We finally want to establish more robust fundamental parameters for this cluster.
    }
   {We selected high S/N high resolution spectra of 7 stars of the Open Cluster NGC~6475 from
    the ESO database covering the T$_{\rm eff}$ range 4500-10000 K and of luminosity
    class V (dwarf) and III (giants). 
    We determined the chemical abundances of several elements. For hot stars 
    (T$_{\rm eff}$$>$9000 K) we applied the Balmer Lines fitting method
    to obtained atmospheric parameters. For cool stars (T$_{\rm eff}$$<$6500
    K) we used the FeI/II abundance equilibrium method. For the two groups
    of stars the use of different line-lists was mandatory. LTE approximation
    was used. For elements affected by NLTE deviation (C,N,O,Na,Mg) corrections
    were applied. The abundance of many elements were obtained from the
    measurement of the equivalent width of spectral lines. For those elements
    for which only blended lines were available (O, He) comparison of real 
    spectrum with synthetic ones was used. Hyperfine structure was taken in 
    account for V and Ba.
    }
   {First of all we showed that the two methodologies we used give abundances which 
    are in agreement within the errors. This implies that no appreciable relative 
    systematic effects are present for the derived chemical content of cool and hot stars.
    On the other hand giants stars show clear chemical peculiarities with
    respect the dwarf concerning light elements (up to Si) and maybe Ba.
    This fact can be explained as an evolutionary effect.
    Then, having a new estimation of the metallicity for the cluster
    ([Fe/H]=+0.03$\pm$0.02, [$\alpha$/Fe]=-0.06$\pm$0.02) 
    we fitted suitable isochrones to the CMD of the cluster obtaining
    the basic parameters (E(B-V)=0.08$\pm$0.02, (m-M)$_0$=7.65$\pm$0.05, Age=200$\pm$50). 
    }
   {}


   \maketitle
%
%________________________________________________________________
                               
\section{Introduction}

Metal abundance determination of individual stars in Galactic open clusters
brings useful information on the formation and chemical evolution of a cluster
itself, on the importance of mixing and rotation in the surface chemical properties
of each star and, finally, allow to determine solid estimate of cluster
bulk properties as distance, reddening, age and age spread.
Very detailed studies are now available, like the ones
on Coma Berenices \citep{GM08b}, the Pleiades \citep{GM08a},
and Praesepe \citep{Fo07,Fo08}, to make a few examples.\\
The ESO archive contains high resolution spectroscopic data of many
stars in Galactic clusters which have been taken for different
purposes, but that have never been fully investigated.
We searched the archive and found several interesting cases, and
present here a detailed analysis of individual high resolution
spectra of stars in the open cluster NGC~6475.
This cluster was studied before several times
(\citealt{Pr96}, \citealt{Fo07}, \citealt{Me93}, \citealt{Ka05}), 
and it turned out to have a reddening in the range 
0.06-0.10 mag and a distance modulus not well constrained but in the range 
(m-M)$_{\rm V}$=7.0-7.7 mag.  Age was found to be in the range 170-220 Myr,
and the metallicity is slightly super-solar ([Fe/H]=+0.14 according to \citealt{Se03}).
The cluster possesses both hot stars of spectral type around B, 
and evolved red giants of spectral type about K.\\
However a throughout chemical investigation of both cool and hot
stars together has never been performed.\\
In this paper we describe in deep details the techniques in use
to infer abundances for a variety of elements in hot and cool stars (both
dwarf and evolved), giving special emphasis to the comparison of the results for elements
belonging to the same group ($\alpha$, iron peak or light elements).
The results are then used to revise the fundamental parameters of the cluster.\\
The layout of this paper is as follows.
In Sect.~2 we describe the observation material, {\bf while in Sect.~3 the
build-up of suitable line-lists is illustrated.
In Sects. 4 and 5 we discuss the determination of atmospheric parameters for cool and hot stars
respectively, and the abundance determination is analyzed in Sect.~6.}
A brief re-visitation of the cluster parameters is finally given in Sect.~7,
while the paper conclusions are highlighted in Sect.~8.

\begin{table*}
\caption{Basic parameters for the observed stars.}            
\label{t1}      
\centering
\begin{tabular}{lcccccccccc}        
\hline\hline                 
ID & $\alpha$(h:m:s) & $\delta$($^{\rm o}$:$^{\prime}$:$^{\prime}$$^{\prime}$) & V(mag) & B-V(mag) &\
Sp.T. & RV$_{\rm H}$(km/s) & v$_{\rm e}$sini(km/s) & T$_{\rm {eff}}$(K) & log(g)(dex) & v$_{\rm t}$(km/s)\\    
\hline             
HD 162679 & 17:53:45.9 & -34:47:29 &  7.16 & 0.03 & B9V   & -17.06 & 37 & 9962 & 3.46 & 0.40 \\
HD 162817 & 17:54:27.1 & -34:28:00 &  6.11 & 0.04 & B9.5V & -15.66 & 65 & 9651 & 3.29 & 0.85 \\
HD 162391 & 17:52:19.8 & -34:25:00 &  5.85 & 1.10 & G8III & -14.51 &  - & 4800 & 1.60 & 1.83 \\
HD 162587 & 17:53:23.5 & -34:53:42 &  5.60 & 1.09 & K3III & -17.18 &  - & 4850 & 2.10 & 1.65 \\
JJ10      & 17:53:54.2 & -34:46:08 & 12.53 & 0.74 & K0V   & -14.87 &  - & 5700 & 4.30 & 1.10 \\
JJ22      & 17:53:08.9 & -34:45:52 & 11.11 & 0.50 & F5-G0V& -14.96 &  - & 6300 & 4.05 & 1.25 \\
JJ8       & 17:54:09.7 & -34:53:13 & 13.30 & 0.87 & K5V   & -14.83 &  - & 5400 & 4.50 & 0.85 \\
\hline                                   
\end{tabular}
\end{table*}

\section{Observations and Data Reduction}

Observations of stars in the field of the open cluster NGC~6475
were carried out during August 2001
in the context of the ESO DDT Program ID 266.D-5655 \citep{Ba03}.
This program had the aim of observing spectroscopically a large sample of field 
and cluster stars from ultraviolet (UV) to near infrared (NIR).\\
Observations were performed with UVES 
on board UT2(Kueyen) telescope in Paranal.
Spectra of 32 stars candidate cluster members (selected
from WEDBA database\footnote{http://www.univie.ac.at/webda/})
were obtained using the DIC1(346+580) and DIC2 (437+860) settings
with a 0.5$^{\prime\prime}$ slit width.
The spectra cover the wavelength range 
3000-10000 \AA\ with a mean resolution of 80000.\\
Data were reduced using the UVES pipeline \citep{Ba00}
where raw data were bias-subtracted, flat-field corrected,
extracted using the average extraction method
and wavelength calibrated. Sky subtraction was applied.
Echelle orders were flux calibrated using the 
master response curve of the instrument, and 
an atmospheric extinction correction was applied. 
Finally, the orders were merged to obtain a 1D spectrum.
All the reduced spectra can be downloaded from the  UVES POP
web interface\footnote{http://www.sc.eso.org/santiago/uvespop/index.html}.\\
For our purposes we selected stars having low rotation (v$_{\rm R}<$100 km/s)
allowing the measurement of equivalent width (EQW)
of spectral lines.\\
Our choice left us with 7 members, {\bf covering the spectral type range K5-B9
and of luminosity class V (dwarf) and III (giants).}
S/N of their spectra vary from star to star and it is function of the wavelength,
but it is greater than 200 in the worst case.\\
Membership was checked by radial velocity measurement that was 
obtained using the {\it fxcor} IRAF, which cross-correlates the observed 
spectrum with a template having known radial velocity. 
As templates we used two synthetic spectra, the former for cool stars and
calculated for a typical Sun-like dwarf (T$_{\rm {eff}}$=5777, 
log(g)=4.44, v$_{\rm t}$=0.80 km/s, [Fe/H]=0.0),
and the latter for hot stars and calculated for a typical A0V dwarf 
(T$_{\rm {eff}}$=9500, log(g)=4.00, v$_{\rm t}$=1.00 km/s, [Fe/H]=0.0).
Spectra were calculated using the 2.73 version of SPECTRUM,  the local
thermodynamical equilibrium spectral synthesis program freely distributed by
Richard O. Gray\footnote{See http://www.phys.appstate.edu/spectrum/spectrum.html for more details.}.\\
The error in radial velocity - derived from {\it fxcor} routine - is less
than 1 km/s. According to the measured RV$_{\rm H}$ and to the typical 
velocity dispersion of Open Clusters ($\sim$1-2 km/s) all the selected 
stars turn out to be members.
Finally, for the abundance analysis, each spectrum was shifted
to rest-frame velocity and continuum-normalized.\\
In Table~\ref{t1} we report the basic parameters of the selected stars:
the identification (ID), the coordinates ($\alpha$, $\delta$), V magnitude, B-V color,
Spectral Type (Sp.T.), heliocentric radial velocity (RV$_{\rm H}$), rotational velocity for the
hot stars (v$_{\rm e}$sini), effective temperature T$_{\rm {eff}}$, gravity log(g), and microturbolence
velocity v$_{\rm t}$ (for determination of rotational velocity and atmospheric parameters
see Sec. 4 and 5).\\
According to SIMBAD database\footnote{http://simbad.u-strasbg.fr/simbad/},
three of them (HD~162679, HD~162391, and HD~162587) are binaries, but no
evidence of a double spectrum was found (no double peak in the
cross-correlation function in the radial velocity determination). We conclude that 
the secondary component must have negligible influence on the spectral 
energy distribution, and so no effects on our spectroscopic analysis are
expected.\\
We do not have any information on the other 4 stars, but 3 of
them (JJ8, JJ10, JJ22, see Fig.~\ref{f3}) lie in the single star MS region,
well detached from the binary sequence. In any case all of them show no
evidence of a double spectrum or double peak in the cross-correlation function 
when the radial velocity determination was performed. Also in this case
we conclude that the secondary component, if present, has negligible
influence on the spectral energy distribution, and so no effects on our 
spectroscopic analysis are expected.

\begin{figure*}
\centering
\includegraphics[width=8cm]{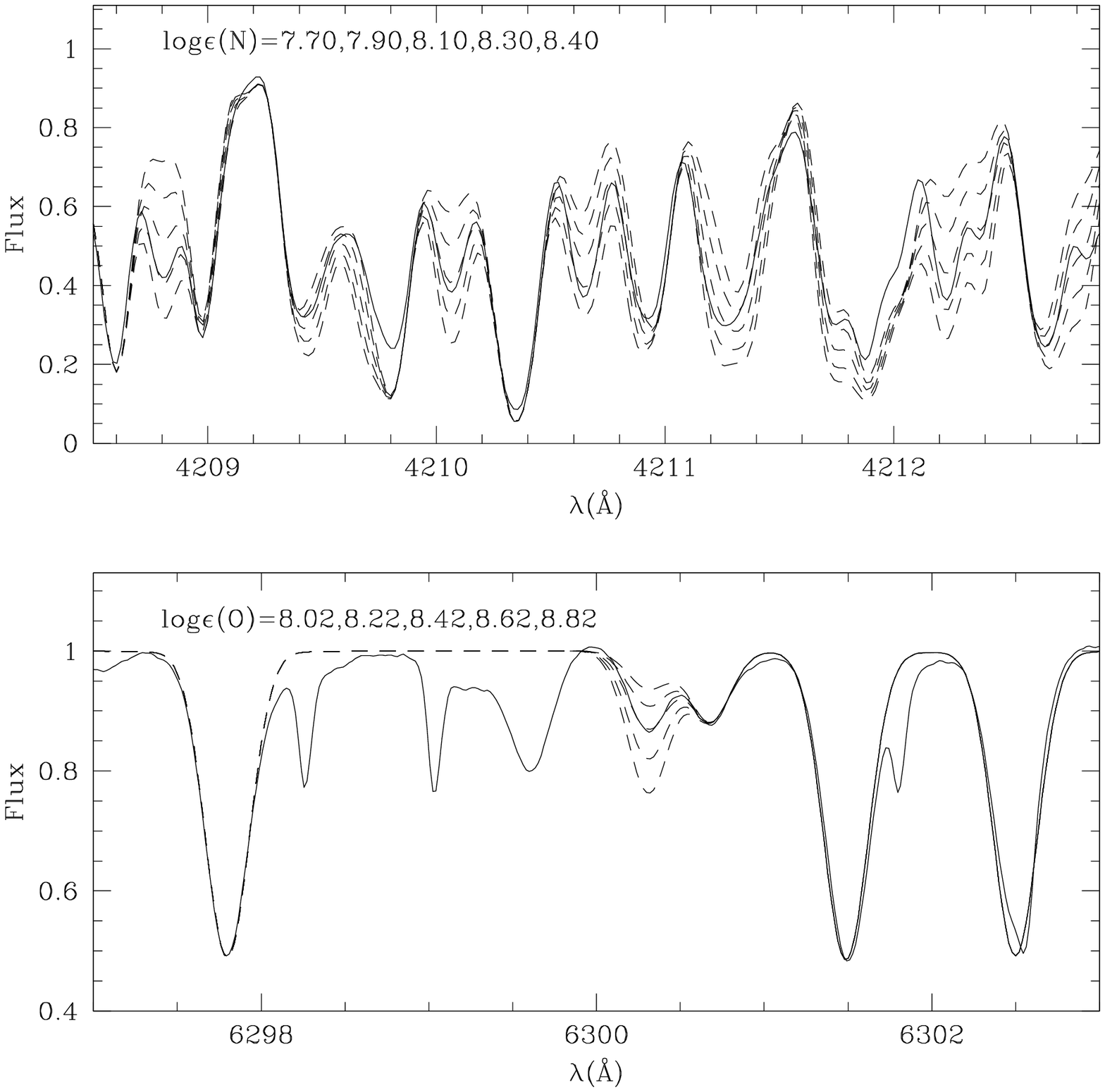}
\includegraphics[width=8cm]{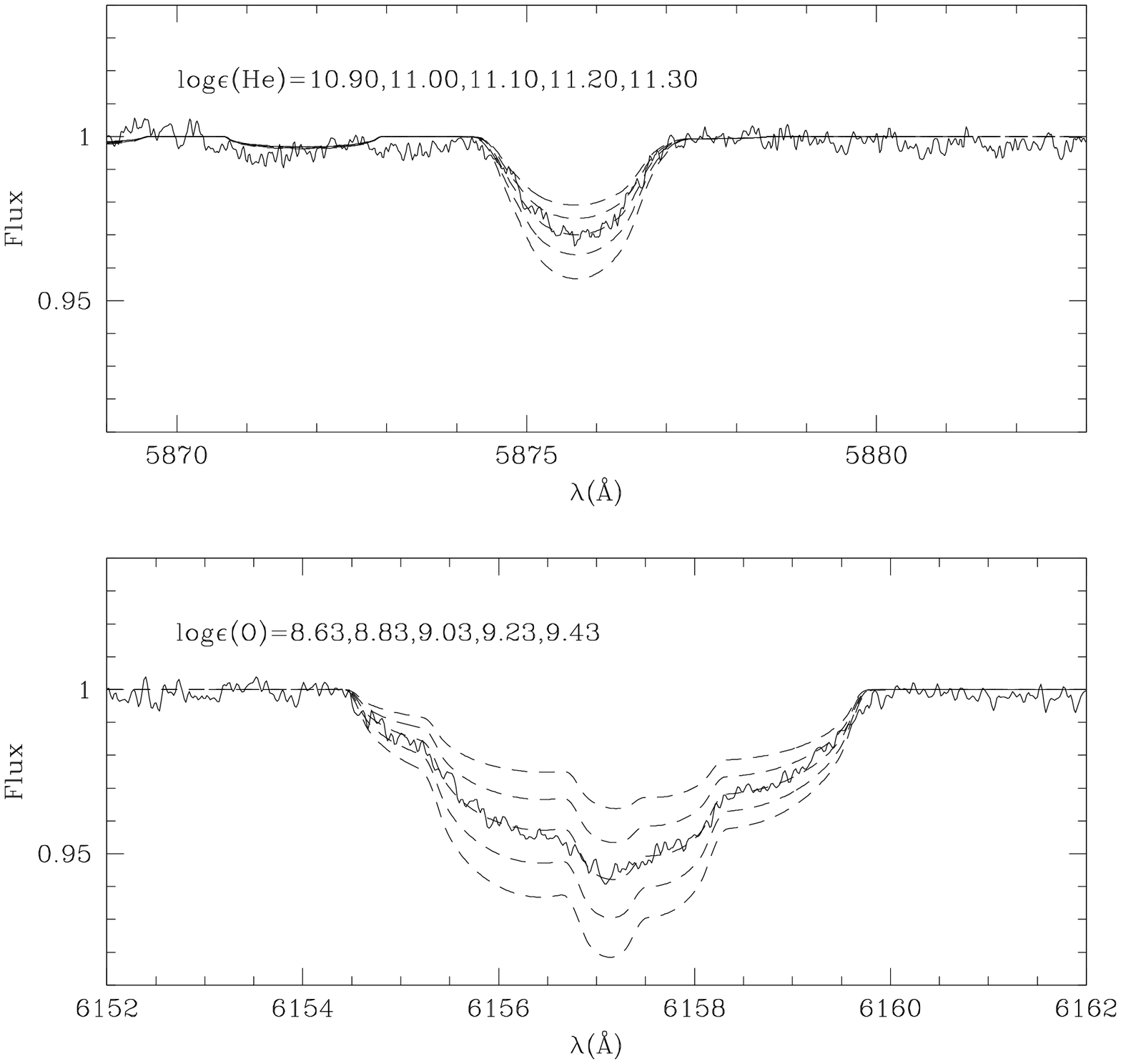}
\caption{Example of the spectral synthesis method used to derive
N and O for cool stars (applied to \#HD~162391, left panel), and He 
and O for hot stars (applied to \#HD~162817, right panel). 
Continuous lines are the observed spectra, while dashed lines are
synthetic spectra. Abundances used in the synthesis are indicated. For N abundace 
determination spectral regions of 0.4 \AA\ width centered at 4209.8 and 4212.0 \AA\ 
were rejected because of the very poor reproduction of the observed spectrum.
The reason is that some lines are not identified in the solar spectrum.
We tentativelly attributed these lines to Fe, which appears to be a bad
choice, so we did not include the relative spectral regions in the fitting.}
\label{f0}
\end{figure*}

\section{The line-lists}

{\bf Early (Hot) and late (cool) type} stars in our sample have few if not common
spectral lines. For this reason we had to build two different
line-lists for {\bf the purpose of measuring their abundances}.
Abundances for most of the elements can be obtained by the EQW method
in both cases.\\
Line-list for cool stars was initially taken from \citet{Gr03}.
log(gf) parameters were then re-determined for each element by a solar-inverse analysis
in order to remove the scatter in abundance with respect to the mean value.
We measured the equivalent widths from the NOAO solar spectrum \citep{Ku84}
by Gaussian fitting (and using a voight profile for the strongest 
lines) and derived the abundances using a model atmosphere 
for  the canonical solar parameters: T$_{\rm {eff}}$=5777 K, 
log(g)=4.44, v$_{\rm t}$=0.80 km/s, log$\epsilon$(Fe)=7.50 dex.\\
Solar abundances we obtained from this line-list are reported 
in Table~\ref{t2} and compared with \citet{GS98}.\\
For some elements (Al and Nd) the agreement was not good
($\Delta$A$\geq$0.2 dex); in this case we redetermined log(gf) values
by averaging the  data obtained from VALD \& NIST\footnote{VALD and NIST database can 
be found at http://vald.astro.univie.ac.at/ and http://physics.nist.gov/PhysRefData/ASD/lines\_form.html} 
atomic parameters databases. Abundances from the new log(gf) parameters 
showed a better agreement ($\Delta$A$<$0.1 dex), and again they were 
adjusted by solar-inverse analysis.
For N and O abundance we were forced to apply the spectral synthesis method
because these lines are affected by severe blending with other spectral features.
O abundances were obtained from the 6300 \AA\ forbidden line, while N from
CN features due to the $\Delta$v=-1 band of the violet system (B$^2$$\Sigma$-X$^2$$\Sigma$)
near the band-head at 4215 \AA.
Line-lists for the synthesis of the O line and CN band were taken from
the more complete lines compilation of SPECTRUM\footnote{See 
http://www.phys.appstate.edu/spectrum/spectrum.html and references therein.}.
The list for the CN feature was calibrated on the Sun in
order to match  the synthetic spectrum with the NOAO one. 
In this case a complete calibration procedure would require
also a comparison with the spectrum of a late type and well known
stars (i.e. Arcturus), in order to adjust log(gf) value of those lines not
present in the Sun, but affecting cooler stars such as HD~162391 and
HD~162587. This is a very long process we will do in future papers. 
For the moment we can say that even without this adjustment 
the agreement between synthetic and observed spectra for HD~16391 
and HD~162587 is satisfactory (see Fig.~\ref{f0}).\\
{\bf Another effect to consider is the hyperfine structure of the odd-numbered elements
such as Sc, V, Co, Cu, Y and Ba. These elements possess non-zero nuclear spin, which result in
considerable splitting of the lines into hyperfine components. According
to \citet{Mc94}, in the case of weak lines this split affects only the shape 
of the line and not the EQW, leaving the derived abundance un-altered.
On the other hand strong lines are desaturated, increasing their EQW,
whose final effect is a anomalously greater abundance for the element if no
correction is applied.\\
For Sc, Co, Cu, and Y we considered only weak lines (EQW$\simeq$ 50
m\AA\ or lower according to \citet{Mc94}). For V and Ba only
strong lines were available, and we derived their abundances from 6274,
6285 and 6812 \AA\ features for V, and from 5853 and 6496 \AA\ features
for Ba respectively. In both cases spectral synthesis was applied and the hyperfine
components were taken from \citet{Mc94} and \citet{Mc98}. 
Ba lines have also isotopic components, and for the isotopic ratios
we assumed the solar values as described in \citet{Mc98}.}\\
As a results of this procedure all our abundances agree with \citet{GS98}
within 0.1 dex except for BaII for which a difference of $\sim$0.20 dex was
found (see Table~\ref{t2}).\\

Line-list for hot stars was taken from VALD \citep{Ku00} database for a typical
A0V solar metallicity star.
Then log(gf) parameters were then re-determined by a inverse analysis
measuring the equivalent widths on a reference spectrum: the Vega spectrum
was chosen for this purpose.
We measured EQWs on two Vega spectra, the first (4100-6800 \AA\ wavelength
range) obtained from Elodie database \citep{Mo04}, and the second 
(3900-8800 \AA\ wavelength range) from \citet{Ta07}. Since Vega is a rotating star,
EQWs were obtained from direct integration of spectral features.
In the common wavelength range (4100-6800 \AA) the measured EQWs were
averaged.
The atmospheric parameters for Vega were obtained in the following way.
As for T$_{\rm eff}$, Vega is primary standard because both angular
diameter and bolometric flux are available. 
From the literature
we obtain the following information: {\it f}=29.83$\pm$1.20$\times$10$^{-9}$ W/m$^2$, 
$\theta$=3.223$\pm$0.008 mas. Hence,  we infer an effective temperature
of 9640$\pm$100 K.\\
Since a direct estimate of Vega gravity is not available, we determined
log(g) using H Balmer lines fitting. We obtained log(g)=3.97$\pm$0.05.
The microturbulence velocity was obtained by minimizing the 
slope of the abundances obtained from FeI lines vs. EQW. We obtained
v$_{\rm t}$=1.02$\pm$0.05 km/s and log$\epsilon$(Fe)=7.02 dex.
The iron content normalized to the Sun results [Fe/H]$\sim$-0.5,
in agreement with previous determinations \citep{Qi01}.
Vega abundances we obtained are reported 
in Table~\ref{t2} and compared with \citet{Qi01} and with the Sun.\\
Also in this case for some elements (He,O) we were forced to apply the
spectral synthesis method which compares spectral features to
synthetic spectra having different abundances of the studied element.
In fact, spectral features of He (5875 \AA) and O (6156-6158 \AA)
are composed by several blended lines of the element.
Suitable line-lists for the synthetic spectrum calcolation were
taken from VALD \& NIST databases and the log(gf) values averaged.\\
It was not possible to derive He abundance for Vega because the target
spectral line (5875 \AA) was too contaminated by telluric lines.\\ 

All the analysis (both for the Sun and Vega, and for the target stars) was
performed using the 2007 version of MOOG \citep{Sn73}
under LTE approximation coupled with ATLAS9 model atmospheres \citep{Ku92}.\\
Spectral features affected by telluric contamination were rejected.\\
C, O, Na and Mg abundances were corrected for NLTE effects when necessary 
as described in the {\bf next} Sections, while N content for Vega and the hot
stars in our sample was corrected as described in Sec. 6.\\

\section{Atmospheric Parameters for Cool stars}

For cool stars (HD~162391, HD~162587, JJ10, JJ22, JJ8) the 
classical method for obtaining spectroscopic atmospheric
parameters is to use the abundances from EQW of FeI/FeII lines.
Initial estimates of the atmospheric parameter T$_{\rm eff}$ 
were obtained from the Sp.T. reported in Table~\ref{t1} according
to the relations given by \citet{SK81}.
We then adjusted the effective temperature by removing any trend in the relation
between the abundance obtained from Fe I lines and the excitation potential.
At the same time, the input log(g) values were {\bf set in order} to satisfy the
ionization equilibrium of FeI and FeII until the relation:
\ \\
\begin{center}
${\rm log\epsilon(FeII)_{\odot}-log\epsilon(FeI)_{\odot}=log\epsilon(FeII)_{\star}-log\epsilon(FeI)_{\star}}$
\end{center}
\ \\
was accomplished.\\
Finally, the microturbulence velocity was obtained by removing any slope 
in the relation between the abundance from FeI lines and the reduced EQW.
Typical internal errors for this method and the S/N of our spectra 
are: $\Delta$T$_{\rm eff}$$\sim$30-40 K, $\Delta$log(g)$\sim$0.1, 
$\Delta$v$_{\rm t}$$\sim$0.05 km/s (see \citet{Ma08}).
Adopted values for the atmospheric parameters are reported in
Table~\ref{t1}.\\
Abundances for most of the elements were obtained from EQW measurements.
All the cool stars do not show evidence of rotation, and for this reason EQWs were 
obtained from Gaussian fitting of spectral features.\\
For N and O abundance we were forced to apply the spectral synthesis method
as said in the previous section. See Fig.\ref{f0} for an example of the
synthesis for N and O applied to \#HD~162391.\\ 
In addition we applied the spectral synthesis method to the forbidden blended
CI line at 8727 \AA\ to complete the C abundances obtained from
EQW of the other unblended C features.\\
CI-8727 is formed strictly in LTE mode at odd with the 
other features ($\lambda$=4775,5052,5380 \AA), so our aim was to compare 
the abundances obtained from the two set of lines to evaluate possible NLTE effects.
We did not find  significant differences ($\Delta$[C/H]$<$0.05, in 
agreement with \citealt{Ta05}), and therefore we did not apply any NLTE corrections 
to the derived abundance for this element.\\
Na content was derived from the 5682-5688 and 6154-6160 \AA\ doublets, while
Mg one from 5711,6318,6319 high exitacion lines. Both Na and Mg lines are known to be
affected by a not negligible amount of NLTE effect. {\bf For this reason abundances obtained 
in LTE approximation have been corrected according to \citet{Gr99}. Each line was treated 
separately because the amount of the correction is a function of the EQW, besides
the atmospheric parameters of the star.
In Table~\ref{t2} and \ref{t3} we report only the mean LTE and NLTE abundances
for the Sun and for our targets respectivelly.}\\
Unfortunately incompleted sources of NLTE correction for Al abundances
are available in literature, which provides corrections for
dwarf or sub-giants stars in the range K0-F5 (\citealt{Ba97} and \citealt{Ge04}).
However, according to \citet{As05} NLTE corrections for Al are important
for  metal poor stars ([Fe/H]$<$-0.5) and less important at
decreasing temperature. Also gravity appears to have a (small) influence on
the correction but no data are available for giant stars. 
Since cool stars in our sample are metal rich and 
the two giants (HD~162391 and HD~162587) are very cool (T$_{\rm eff}$$\sim$4800 K), we estimate
that no NLTE corrections appear necessary for this element.\\ 
 
\begin{figure*}
\centering
\includegraphics[width=14cm]{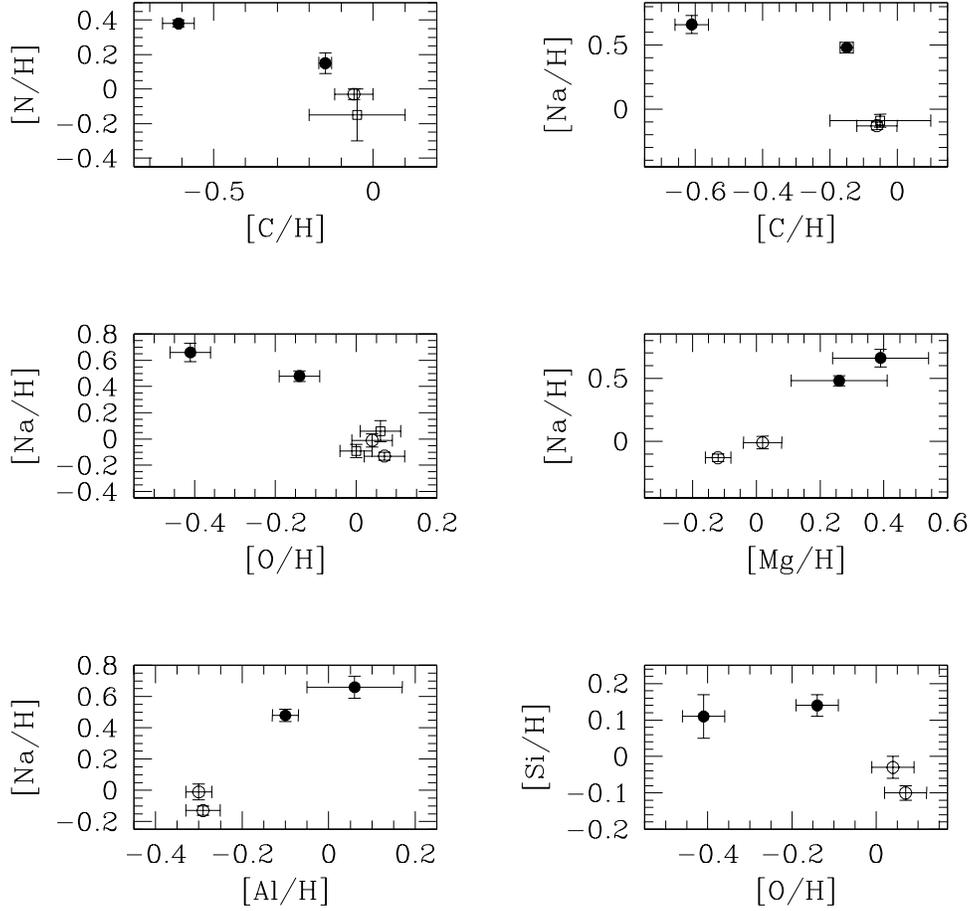}
\caption{The correlations N-C, Na-C, Na-O, Na-Mg, Na-Al, and Si-O for
         our stars. Filled circles are the giants, open ones are the
         cool dwarf, while open squares are the hot dwarfs.}
\label{f1}
\end{figure*}

\begin{table}
\caption{Abundances for the Sun and Vega as obtained from
the line lists used in this work. Abundances for the Sun are
compared with \citet{GS98} (Sun$_{\rm GS98}$), those for 
Vega with \citet{Qi01} (Vega$_{\rm Qi01}$). For some elements
(C, N, O, Na, \& Mg) NLTE correction were applied and both LTE
and NLTE abundances reported. For V and Ba we took into account the
hyperfine structure affecting their lines.}    
\label{t2}      
\centering                          
\begin{tabular}{lcccc}        
\hline\hline 
El. & Sun & Sun$_{\rm GS98}$ & Vega & Vega$_{\rm Qi01}$\\
\hline
HeI      &  -   &10.93 &  -   &  -  \\
CI       & 8.50 & 8.52 & 8.52 & 8.46\\
CI$_{\rm NLTE}$ & - & - & 8.57 & -  \\
NI       & 7.95 & 7.92 & 7.98 & 8.00 \\
NI$_{\rm NLTE}$ & - & - & 7.58 & - \\
OI       & 8.83 & 8.83 & 8.82 & 9.01\\ 
OI$_{\rm NLTE}$ & - & - & 8.74 & -  \\   
NaI      & 6.37 & 6.33 & 6.63 & 6.45\\   
NaI$_{\rm NLTE}$ & 6.32 & 6.33 & 6.33 &  -  \\
MgI      & 7.54 & 7.58 & 7.14 & 6.81\\  
MgI$_{\rm NLTE}$ & 7.56 & 7.58 & - &  -  \\
AlI      & 6.43 & 6.47 &  -   &  -  \\
SiI      & 7.61 & 7.55 &  -   &  -  \\
CaI      & 6.39 & 6.36 & 5.65 & 5.41\\  
ScII     & 3.12 & 3.17 & 2.56 & 2.33\\ 
TiI      & 4.94 & 5.02 &  -   &  -  \\
TiII     & 4.96 & 5.02 & 4.52 & 4.58\\
VI$_{\rm Hyp}$       & 4.00 & 4.00 &  -   &  -  \\
CrI      & 5.63 & 5.67 &  -   &  -  \\
CrII     & 5.67 & 5.67 & 5.20 & 5.19\\ 
FeI      & 7.50 & 7.50 & 7.03 & 6.94\\
FeII     & 7.51 & 7.50 & 7.01 & 6.93\\
CoI      & 4.85 & 4.92 &  -   &  -  \\
NiI      & 6.28 & 6.25 &  -   &  -  \\ 
CuI      & 4.19 & 4.21 &  -   &  -  \\ 
ZnI      & 4.61 & 4.60 &  -   &  -  \\ 
YII      & 2.24 & 2.24 &  -   &  -  \\ 
BaII$_{\rm Hyp}$     & 2.34 & 2.13 & 1.85 & 0.81\\ 
LaII     & 1.26 & 1.17 &  -   &  -  \\ 
CeII     & 1.53 & 1.58 &  -   &  -  \\ 
NdII     & 1.59 & 1.50 &  -   &  -  \\ 
\hline                                   
\end{tabular}
\end{table}

\section{Atmospheric Parameters for Hot stars}

Atmospheric parameters of hot stars (HD~162679, HD~162817) are routinely
determined by fitting the Balmer lines of observed spectra
with synthetic ones.
For our analysis we used a sets of ATLAS9 model atmospheres 
calculated for solar metallicity (roughly the metallicity
of the cluster as derived by this study).
Starting from these model atmospheres we calculated spectra with Lemke's
version of the LINFOR program (developed originally by Holweger, 
Steffen, and Steenbock at Kiel University).\\ 
To achieve the best fit, we used a
routine  which employs a $\chi^2$ test. The $\sigma$ necessary 
for the calculation of $\chi^2$ is estimated from the noise 
in the continuum regions of the spectra .The fit program normalizes
model spectra {\em and} observed spectra using the same points for the
continuum definition.  We used the Balmer lines H$_\beta$ to H$_{12}$
(excluding H$_\epsilon$ to avoid the CaII~H line) for the
fit.\\
The formal errors given by the routine are $\Delta$T$_{\rm}$=10-20 K,
$\Delta$log(g)=0.005-0.01, which, according to \citet{Mo06} are half of the true
value.
For this reason we assumed as internal uncertainties for our determinations
of: $\Delta$T$_{\rm eff}$=20-40 K and $\Delta$log(g)=0.01-0.02 dex.\\
Microturbulence velocity was obtained  by removing any trend in the
relation between abundance and reduced EQW for FeI and FeII lines
and the typical internal error is 0.1 km/s, obtained as in \citet{Ma08}.\\
The adopted values for the atmospheric parameters are reported in
Table~\ref{t1}.\\

Abundances for most of the elements were obtained from EQW measurements.
The selected hot stars show clear evidence of
rotation.  As a consequence, we refrained to obtain EQWs from a Gaussian fitting,
but used a direct integration of spectral features.\\
On the other hand, for some elements (He,O) we were forced to apply the
spectral synthesis method as explained in Sec. 3. See Fig.\ref{f0} for an example of the
synthesis for He and O applied to \#HD~162817.\\ \\
From spectral synthesis we were able to measure also the projected rotational velocity
of the stars (reported in Table\ref{t1}) by confronting the shape of the spectral
lines with synthetic spectra. The typical  rotational velocity error is 3-5
km/s (obtained by comparing rotational velocities from different lines).\\
Na abundance was derived from the 5889-5895 \AA~ doublet, known to be
affected by strong NLTE effect. NLTE correction for this element
can be large (up to -0.50 dex according to \citealt{Ta03}).
For hot stars in our regime a full discussion of NLTE correction for Na-double
is done in \citet{Ma00}. For our targets a correction of -0.30 dex is required.\\
Also the O triplet at 6156-6158 \AA\ is affected by NLTE and in this 
case we applied corrections by \citet{Ta97}.\\
The NLTE correction for N, necessary for hot stars, is quite a disputed issue 
(see \citealt{Ta92} and \citealt{Le96}) and will be discussed in Sec. 6,
while C abundances were corrected according to \citet{Ta92}.\\ 
No NLTE corrections are available in literature for He or Mg in our T$_{\rm eff}$ regime.
However the Mg abundances we obtained in LTE approximation well agree with those ones
obtained from cool dwarf stars, so we conclude that the NLTE correction is not
necessary in this case.\\ 
For He we simply give the LTE value because
any detailed NLTE treatment is beyond the purpose of the paper. However
in \citet{Vi09} we studied a sample 
of hot Globular Cluster members in a T$_{\rm eff}$-log(g) regime comparable to the 
hot stars of the present paper. We found a He content in very good agreement with the
primordial value for the Universe, which suggest that for objects cooler than 
10000 K NLTE correction for He in not very important.
We leave to a future paper a detailed discussion about this argument.\\

\begin{table*}
\caption{Abundances (log(N$_{\rm el.}$/N$_{\rm H}$)+12) obtained for our stars. The
  absolute (9th column) and referred to the Sun (10th column) averaged 
  abundances were calculated rejecting the two giant stars (see Sec. 6).}    
\label{t3}      
\centering                          
\begin{tabular}{lccccccccc}        
\hline\hline 
El.              & HD~162679     & HD~162817     & HD~162391     & HD~162587     & JJ10          & JJ22          & JJ8           & $<$log$\epsilon({\rm El})$$>$ & $<$[El./H]$>$\\
\hline
HeI              &11.10$\pm$0.05 &11.10$\pm$0.05 &  -            &  -            &  -            &  -            &   -           & 11.10$\pm$0.04 & +0.17$\pm$0.04 \\
CI               & 8.33          &   -           & 7.89$\pm$0.05 & 8.35$\pm$0.02 &  -            &  -            & 8.44$\pm$0.06 &   -            &   -            \\
CI$_{\rm NLTE}$  & 8.45          &   -           &  -            &  -            &  -            &  -            &   -           &  8.44$\pm$0.01 & -0.06$\pm$0.01 \\  
NI               & 8.20          &   -           & 8.33$\pm$0.02 & 8.10$\pm$0.06 & 7.92$\pm$0.04 &  -            & 7.92$\pm$0.03 &  7.92$\pm$0.02 & -0.03$\pm$0.02 \\
NI$_{\rm NLTE}$  & 7.80          &   -           &  -            &  -            &  -            &  -            &   -           &   -            &   -            \\
OI               & 8.97$\pm$0.04 & 9.03$\pm$0.05 & 8.42$\pm$0.05 & 8.69$\pm$0.05 & 8.87$\pm$0.05 &  -            & 8.90$\pm$0.05 &   -            &   -            \\  
OI$_{\rm NLTE}$  & 8.83$\pm$0.04 & 8.89$\pm$0.05 &  -            &  -            &  -            &  -            &   -           &  8.87$\pm$0.02 & +0.04$\pm$0.02 \\
NaI              & 6.53$\pm$0.05 & 6.68$\pm$0.08 & 6.84$\pm$0.07 & 6.73$\pm$0.04 & 6.38$\pm$0.05 &  -            & 6.25$\pm$0.03 &   -            &   -            \\
NaI$_{\rm NLTE}$ & 6.23$\pm$0.05 & 6.38$\pm$0.08 & 6.98$\pm$0.07 & 6.80$\pm$0.04 & 6.31$\pm$0.05 &  -            & 6.19$\pm$0.03 &  6.24$\pm$0.05 & -0.08$\pm$0.05 \\
MgI              & 7.43          & 7.61$\pm$0.09 & 7.59          & 7.60          & 7.54$\pm$0.06 & 7.45$\pm$0.03 & 7.43$\pm$0.04 &   -            &   -            \\
MgI$_{\rm NLTE}$ &  -            &   -           & 7.95          & 7.82          & 7.58$\pm$0.06 & 7.56$\pm$0.03 & 7.44$\pm$0.04 &  7.53$\pm$0.04 & -0.03$\pm$0.04 \\
AlI              &  -            &   -           & 6.49$\pm$0.11 & 6.33$\pm$0.03 & 6.13$\pm$0.03 &  -            & 6.14$\pm$0.04 &  6.13$\pm$0.01 & -0.30$\pm$0.01 \\
SiI              &  -            &   -           & 7.72$\pm$0.06 & 7.75$\pm$0.03 & 7.58$\pm$0.03 & 7.52$\pm$0.04 & 7.51$\pm$0.02 &  7.53$\pm$0.02 & -0.08$\pm$0.02 \\
CaI              &  -            &   -           & 6.40$\pm$0.10 & 6.41$\pm$0.06 & 6.44$\pm$0.04 & 6.39$\pm$0.05 & 6.35$\pm$0.04 &  6.40$\pm$0.01 & +0.01$\pm$0.01 \\
ScII             &  -            & 3.02          &  -            &  -            & 3.05$\pm$0.04 & 3.04$\pm$0.05 & 3.07$\pm$0.06 &  3.05$\pm$0.01 & -0.07$\pm$0.01 \\
TiI              &  -            &   -           & 4.91$\pm$0.05 & 4.98$\pm$0.03 & 4.90$\pm$0.03 & 4.92$\pm$0.05 & 4.96$\pm$0.02 &  4.95$\pm$0.02 & +0.01$\pm$0.02 \\
TiII             & 4.87$\pm$0.04 & 4.95$\pm$0.04 & 4.99$\pm$0.03 & 4.93$\pm$0.04 & 4.89$\pm$0.02 & 4.87$\pm$0.04 & 4.91$\pm$0.05 &  4.91$\pm$0.02 & -0.05$\pm$0.02 \\
VI               &  -            &   -           & 3.83$\pm$0.01 & 3.96$\pm$0.01 & 3.92$\pm$0.02 &  -            & 4.03$\pm$0.01 &  3.94$\pm$0.04 & -0.06$\pm$0.04 \\
CrI              &  -            &   -           & 5.68$\pm$0.06 & 5.72$\pm$0.03 & 5.69$\pm$0.02 & 5.74$\pm$0.05 & 5.65$\pm$0.02 &  5.68$\pm$0.02 & +0.05$\pm$0.02 \\
CrII             & 5.72$\pm$0.05 & 5.82$\pm$0.07 & 5.72$\pm$0.11 & 5.81$\pm$0.06 & 5.69$\pm$0.04 & 5.66$\pm$0.04 & 5.59$\pm$0.02 &  5.65$\pm$0.04 & -0.02$\pm$0.04 \\
FeI              & 7.46$\pm$0.08 & 7.55$\pm$0.05 & 7.52$\pm$0.01 & 7.56$\pm$0.01 & 7.54$\pm$0.01 & 7.51$\pm$0.01 & 7.47$\pm$0.05 &  7.53$\pm$0.02 & +0.03$\pm$0.02 \\
FeII             & 7.47$\pm$0.03 & 7.62$\pm$0.03 &  -            &  -            &  -            &  -            &  -            &  7.54$\pm$0.07 & +0.03$\pm$0.07 \\
CoI              &  -            &   -           & 4.87$\pm$0.06 & 4.95$\pm$0.06 & 4.74$\pm$0.03 &  -            & 4.79$\pm$0.02 &  4.79$\pm$0.05 & -0.06$\pm$0.05 \\
NiI              &  -            &   -           & 6.28$\pm$0.04 & 6.29$\pm$0.02 & 6.21$\pm$0.02 & 6.25$\pm$0.02 & 6.19$\pm$0.01 &  6.22$\pm$0.02 & -0.06$\pm$0.02 \\
CuI              &  -            &   -           &  -            &  -            & 3.98          &  -            & 4.06$\pm$0.05 &  4.05$\pm$0.05 & -0.14$\pm$0.05 \\
ZnI              &  -            &   -           & 4.63          & 4.48          & 4.46$\pm$0.02 &  -            & 4.48          &  4.46$\pm$0.05 & -0.15$\pm$0.05 \\
YII              &  -            &   -           &  -            & 2.40          & 2.27$\pm$0.03 &  -            & 2.36$\pm$0.05 &  2.30$\pm$0.05 & +0.06$\pm$0.05 \\
BaII             &  -            &   -           & 2.60$\pm$0.05 & 2.65$\pm$0.02 & 2.46$\pm$0.01 &  -            & 2.42$\pm$0.01 &  2.44$\pm$0.02 & +0.13$\pm$0.02 \\
LaII             &  -            &   -           &  -            &  -            &  -            &  -            & 1.39          &  1.39          & +0.13          \\
CeII             &  -            &   -           &  -            & 1.59          &  -            &  -            & 1.72$\pm$0.04 &  1.71$\pm$0.04 & +0.18$\pm$0.04 \\
NdII             &  -            &   -           &  -            &  -            &  -            &  -            & 1.95          &  1.95          & +0.31          \\
\hline                                   
\end{tabular}
\end{table*}

\section{Results of the spectroscopic analysis}

Chemical abundances we obtained are summarized in Table~\ref{t3}.
Table~\ref{t3} demonstrates that there is a good agreement between  
hot and cool dwarf stars as far as the abundances of common elements are concerned,
being the mean abundances of the two groups in agreement within 0.10-0.15 dex
in the worst case.\\
{\bf Based on this good agreement, we can conclude that HD~162679 and HD~162817 are
normal A/B objects, not affected by deviation of the superficial abundances,
which affects peculiar (Ap/Bp) stars. 
This results agrees with \citet{Fol07}, where authors find a
normal solar chemical composition, {\bf within the errors}, for HD~162817. 
On the other hand \citet{Fol07} showed that the other NGC~6475 A/B stars in
their sample  have peculiar composition and present a huge variation of
solar scaled abundances of many elements (from C up to Ba) with respect the
mean chemical content of the cluster as traced by cool dwarf or giant stars. 
This variation can reach the value of $\pm$2.0 dex, depending on the element.
Our variation, if present, is lower than 0.10-0.15 dex for all the element we
studied. Because of this we confirm that HD~162679 and HD~162817 do not
present surface abundance peculiarities of Ap/Bp stars.
This result can be easly explained by the high rotation of HD~162679 and
HD~162817 (v$_{\rm e}$sini$>$35 km/s), which, according to \citet{He03}, mixes the
stellar envelope inhibiting diffusion processes which cause chemical anomalies.\\
{\it Based on this fact we can affirm also that no appreciable differential
systematic errors affect the two methods used for abundance analysis, which
was one of the main aims of this paper. In any case differential systematic errors, if
present, are lower than 0.10-0.15 dex for all the chemical abundance we
determined.}.}\\

About the consistence of our atmospheric parameters for hot stars, \citet{Fol07} find for HD~162817 a larger 
temperature ($\sim$300 K) and  gravity ($\sim$0.1 dex), however in agreement
with our values within 1$\sigma$. Only rotational velocity is out of the
3$\sigma$ limits. This is not critical for our results, except in
the case of He and O abuncances, where spectral synthesis was used. 
However Fig.~\ref{f0} shows that the adopted v$_{\rm e}$sini value 
well reproduces the observed spectrum.\\

A second interesting result concerns light elements (from C to Si), that show
clear trends (as we can see in Fig.~\ref{f1}) defined by the abundances of the
two giant (filled circles).
Those stars turned out to be more C/O-poor and N/Na-rich with respect to the
dwarfs (open circles and squares).\\
These trends or correlations are similar to those found for Globular Clusters (GC),
where they seem to be primordial and due to the different composition of the interstellar
medium from which stars of different ages were formed (see \citealt{Gr04} for extensive
references) with C/O-poor and N/Na-rich stars representing the younger generation.\\
In our case trends are fully explained as an evolutionary effect due to the migration
of light elements produced by the H-burning cycle from deeper regions up to the
photosphere after stars have left the MS.\\
However the two phenomena can be correlated, because in Globular Clusters
the younger generation of C/O-poor and N/Na-rich stars is thought to be born
from material polluted by ejecta of the young massive stars belonging to the older generation. 
The favorite classes of candidate polluters are three: fast-rotating massive
main-sequence stars \citep{De07}, intermediate-mass AGB stars \citep{Da02},
and primordial population III stars \citep{Ch07}.\\
In these stars the original superficial abundance of light elements is altered
by mixing phenomena concerning products of the H-burning process at high temperature 
(\citealt{La93}, \citealt{Pr07}) like C,N,O, and Na, as is the
case of giant stars studied in this paper.\\
In this picture evolved stars in NGC~6475 could simply be the present day version
of those AGB intermediate-mass polluter stars present during the first millions of years of
a GC lifetime that caused the correlation we see nowaday, and that are now disappeared.\\
{\bf According to Table~\ref{t3} giants present also a overabundance of Ba of $\sim$0.2 dex with respect 
the dwarf, but this is more difficul to interpretate for us also if some
evolutionary mixing phenomenon cannot be ruled out}.\\
The last two columns of Table~\ref{t3} report the 'absolute' and 'refered to the
Sun' averaged abundances of the cluster as obtained from from out 7
stars. Light element {\bf and Ba} values were calculated 
rejecting the two giant stars. {\bf Also N abundance of \#HD~162679 was rejected 
in the calculation of the mean N content because no {\it a priori} NLTE correction 
could be applied (see the following discussion).\\
Averaged abundances were calculated using the weighted mean,
where the weight {\it w} is obtained from the abundance error $\sigma$ as {\it w}=$\sigma^{-2}$. For
some elements we could not obtain the error, and we simply assumed $\sigma$=0.15, which
is our upper limits for the error on abundance when a single line is used.\\
Error on the mean abundance is the r.m.s. divided by the root square of the
number of stars used for the calculation, except for N and He, were this
procedure gives unreliably low errors.
For those elements error on the mean was assumed to be the mean error on the
single star, divided by the root square of the number of stars}.\\

\begin{figure}
\centering
\includegraphics[width=8cm]{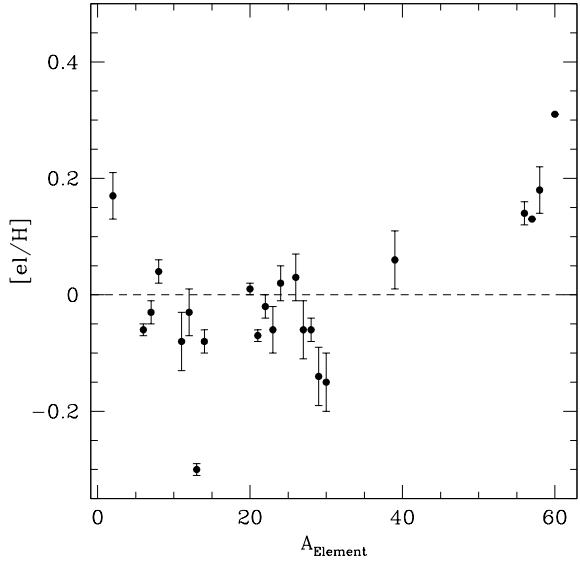}
\caption{Abundances of the cluster with respect of the Sun.}
\label{f2}
\end{figure}

Metallicity of the cluster turns of to be solar or almost solar for most
of the elements, especially for iron peak and $\alpha$ (C,O,Mg,Si,Ca,Ti).
In particular the cluster has:
\ \\
\begin{center}
${\rm [Fe/H]=+0.03\pm0.02\ dex},\ {\rm [\alpha/Fe]=-0.06\pm0.02\ dex}$
\end{center}
\ \\
In spite of this some elements deviate considerably from the solar composition.
He turns out to be supersolar by 0.17 dex. This translates in:
\ \\
\begin{center}
${\rm Y=0.33\pm0.02}$
\end{center}
\ \\
We underline the fact that He was determined in LTE approximation, as
discussed before.\\
Al turns out to be strongly subsolar ($\sim$-0.3 dex).
Cu and Zn result  under-abundant by about -0.15 dex, while La,Ce, and Nd are over-abundant.
Finally s-process elements Y and Ba show {\bf a overabundance of $\sim$0.1 dex}.  
The chemical content of the cluster with respect of the Sun is plotted in
Fig.~\ref{f2}.\\
Some additional discussion is needed for N abundance of the stars HD~162679.
For Vega (that has similar parameters as HD~162679) \citet{Ta92} and
\citet{Le96} reports two different values for the NLTE correction: the former
gives -0.8 dex, while the later -0.4 dex.
Comparing N abundance of HD 162679 with the mean of the dwarf stars (JJ8 and
JJ10) we see that a correction of $\sim$-0.3 dex is required to match the two
values. We conclude that our data suggest
that the right NLTE correction is the one by \citet{Le96}.
So we applied this value to our data, both to HD 162679 and to Vega.\\
Finally,  we want to compare our results with \citet{Se03}, which
determined cluster metallicity from a sample of 30 dwarfs. They give a value
of +0.14$\pm$0.06 for iron content. The agreement with our results is within 1.5
$\sigma$. We attribute this difference to the used methods. \citet{Se03} obtain
temperature and microturbulence from photometry using previously calibrated
colour-T$_{\rm eff}$ and v$_{\rm t}$-T$_{\rm eff}$-log(g) relations. At odd with that,  
our T$_{\rm eff}$ and v$_{\rm t}$ values were obtained directly from
spectra. Therefore it is not surprising that the two results do not perfectly
agree.

\section{NGC 6475 basic parameters revisited}

\begin{figure}
\centering
\includegraphics[width=8cm]{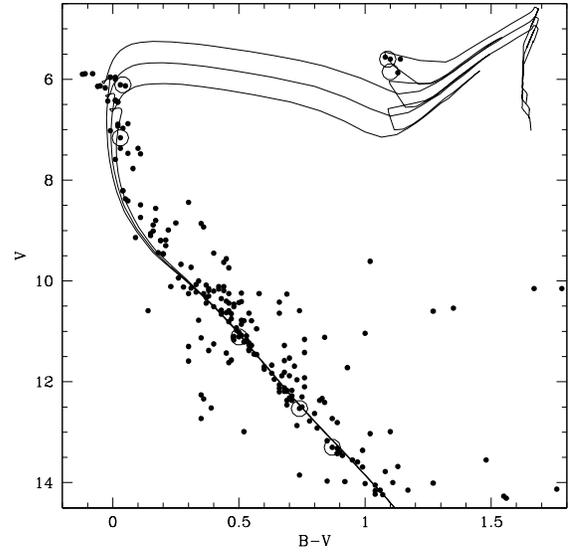}
\caption{CMD of NGC 6475. Isochrones of 150, 200, and 250 Myr are plotted. 
Open circles indicate stars analyzed spectroscopically and confirmed
to be cluster members.}
\label{f3}
\end{figure}

Basing on the detailed chemical analysis detailed in the previous Sections, we
revise here the cluster fundamental parameter using the Padova suite
\citep{Gi00} of stellar models and isochrones. 
Previous determinations of cluster properties are described in \citet{Me93}
and \citet{Ka05}. 
They derive a logarithmic age of 8.35-8.22 respectivelly, which is
compatible with the distance of 280$\pm$26 pc determined by \citet{Ro99}.\\
The cluster metallicity is basically solar, and therefore adopting
Z = 0.019 seems quite an adequate compromise, provided that the amount of
$\alpha-$element under-abundance is almost negligible within the errors.
The procedure is highlighted in Fig.~\ref{f3}, where photometric data taken
from \citet{Pr96} are compared with three solar metallicity
isochrones for ages of 150, 200, and 250 Myr.\\
In general the fit is good, especially for the Main Sequence. For Turn-off
and Red-clump regions (were the cluster population is not well defined)
the fit is not well constrained and stars seems to cover a age range between
150 ({\bf giants}) and 200-250 Myr ({\bf TO stars}). {\bf A reason for this
mismatch could be that giant stars do not have the same superficial
abundance with respect the dwarf due to
evolutionary phenomena, which are not considered in the isochrones}.
The age we derive is therefore 200$\pm$50 Myr.\\
The fit has been obtained shifting the isochrones by E(B-V)=0.08, and
(m-M)$_{\rm V}$=7.65. A reddening of 0.08 mag  is confirmed also by
the two-colors diagram (U-B vs. B-V, not reported here), where the few blue
stars having UBV magnitudes are well fitted by the ZAMS obtained from Padova
isochrones shifted by the previously reported reddening. This value lies
in the middle of the range given by literature (0.06-0.10 mag).\\
Basing on a visual inspection, we estimate errors
of 0.02 and 0.05 mag. on reddening and apparent distance modulus, respectively.
We infer a distance of 300$\pm$10 pc, in agreement
with previous estimates  and which is the value also reported in WEBDA.
We underline the fact that reddening and distance of the cluster are very well
constrained by our study, mainly because we could confirm the former parameter
obtained by the isochrone fitting by the two-colors diagram. On the other hand
age is not well determined because the Turn-Off and Red-Clump regions of the cluster
are not well defined.

\section{Conclusions}

We studied a sample of stars belonging to the Open Cluster NGC~6475 (M7)
covering a wide range in temperature (4500-13000 K) {\bf and gravities (both dwarf and giants)}.
Spectroscopic data were obtained from the UVES POP database, while
photometric ones from WEBDA database.\\
Our aim was to determine abundances of a wide range of elements (from He
to Nd) in different kinds of objects, which require different methodologies
of analysis. Parameters (T$_{\rm eff}$ and log(g)) for cool MS and Giant stars were 
obtained by the FeI/II abundance equilibrium method, while for hot MS stars they were
determined from the shape of H Balmer lines. For the two group of 
objects two sets of line-lists were used, calibrated on the Sun and Vega respectively.
Abundances were obtained in LTE aproximation, but for several 
elements (C,N,O,Na,Mg) NLTE correction from literature was applied
and hyperfine structure was considered for Vanadium {\bf and Barium}.\\
Abundances of common elements in hot and cool MS stars agree very well (within
0.1 dex), allowing to conclude that the methods used for the two kind of objects
are not affected by appreciable relative systematic errors.\\
On the other hand abundances for the two giants stars do not agree 
with the ones obtained for MS as far as light elements and Ba are concerned. 
This is an indication of an evolutionary effect which changes the 
photospherical chemical composition when stars leave the MS.
These stars could be the present day version of those massive polluters
present during the first millions of years of a GC lifetime, that
altered the chemical composition of the intracluster medium from which they
formed as discussed in Sec. 6. This contamination is visible nowaday in the
Na-O anticorrelation found to affect almost all GCs.\\
The Cluster turns out to have solar composition ([Fe/H]=+0.03,[$\alpha$/Fe]=-0.06)
within $\pm$0.1 dex for most of the elements. A overabundance was found
for He and heavy elements (Ba,La,Ce,Nd), while a strong underabundance 
was found for Al.\\
Finally the Cluster parameters were revised. We obtained E(B-V)=0.08,
(m-M)$_{\rm V}$=7.65, and a age of about 200 Myr.

\begin{acknowledgements}
This study made use of ESO, SIMBAD and WEBDA databases.
\end{acknowledgements}

\bibliographystyle{aa}
\bibliography{ms_5.bbl}

\end{document}